\documentclass[prl,twocolumn,showpacs]{revtex4}
\usepackage{graphicx}
\usepackage{float}
\usepackage{amsfonts}
\usepackage{makeidx}
\usepackage{graphicx}
\usepackage{amsmath,amssymb,graphics,epsfig,color,times,bbm}

\begin{document}

\title{Pure Quantum Correlations Between Bright Optical Beams}
\author{Yana Shang, Zhihui Yan}
\author{Xiaojun Jia}
 \email{jiaxj@sxu.edu.cn}
\author{Xiaolong Su}
\author{Changde Xie}
\author{Kunchi Peng }
\affiliation{State Key Laboratory of Quantum Optics and Quantum Optics Devices,\\
Institute of Opto-Electronics, Shanxi University, Taiyuan, 030006,
P. R. China}

\begin{abstract}
The pure quantum correlations totally independent of the classical
coherence of light have been experimentally demonstrated. By
measuring the visibility of the interference fringes and the
correlation variances of amplitude and phase quadratures between a
pair of bright twin optical beams with different frequencies
produced from a non-degenerate optical parametric oscillator, we
found that when classical interference became worse even vanished,
the quadrature quantum correlations were not influenced, completely.
The presented experiment obviously shows the quantum correlations of
light do not necessarily imply the classical coherence.
\end{abstract}


\maketitle

Both quantum and classical correlations of light fields are
extensively applied in optical communication, information processing
and optical measurements. The relationship between quantum and
classical correlations is a widely attended issue,
recently\cite{Sen,Hay,Wolf,Pan,Kas}. Although it has been
theoretically proved that quantum correlations can exist without
accompanying classical correlations\cite{Hay,Kas}, there is no
experiment specially to clarify this arguable problem so far. The
entangled optical modes with quantum correlations of amplitude and
phase quadratures have been successfully utilized in the quantum
information with continuous variables (CV)\cite{Bra} to realize the
unconditional quantum teleportation, quantum dense coding,
entanglement swapping and quantum key distribution\cite
{Fur,Li,Jia,Su2}. The two entangled optical modes used in all
above-mentioned experiments are frequency-degenerate produced from
optical parametric oscillators (OPO) operating below the oscillation
threshold, thus they have also good classical coherence naturally.
In CV quantum information systems the balanced-homodyne-detectors
(BHD) are used for the correlation measurements and the signal
extraction, usually. For directly applying the fundamental wave of
the pump laser to be the local oscillator in BHD the entangled
optical modes with a degenerate frequency are necessary, which
easily make a false intuition to think that the classical coherence
is contained in quantum correlations. The consideration seems
reasonable from the general conception that classical physics is a
special case of quantum physics, so we may think that after
depleting the quantum correlations the classical correlations will
be ultimately retained. It is a well-known fact that the quantum
correlations are much more difficult to be generated and observed
than classical coherence. For example, there is classical coherence
between two optical beams split from a laser always, but there is no
any quantum correlation in them. It is a generally recognized fact
that quantum correlations may vanish when classical correlations
exist. However, no more attentions have been paid to the opposite
problem, especially there is no the experimental study to
demonstrate the existence of quantum correlations without classical
coherence although the theoretical discussion on this problem has
been presented very recently\cite{Kas}.

For the ''bright'' light field the particle effects do not dominate,
the quantum nature of which is mainly characterized by the presence
of quantum noise. The coherent state is a minimum uncertainty state
with equal noise fluctuations in the two quadrature components and
is the closest quantum approximation to the light field generated by
a laser. In an ideal light field of the coherent state all classical
noises vanish and only the quantum noises limited by the minimum
uncertainty of quantum mechanics are retained. The quantum noises of
a coherent state are defined as the quantum noise limit (QNL).

According to quantum mechanics the quantum correlations existing
between spatially separated optical beams can be stronger than that
allowed by classical physics. With bright beams of light the quantum
properties are encapsulated in the noise sidebands around its
carrier frequency. For technical and engineering applications in CV
quantum information systems, we are more interested in correlations
between the amplitude and phase quadrature fluctuations of the
spatially separated beams, recorded by separated detectors. The
quadrature correlations of optical beams can be quantified by
measuring the sum ($V\widehat{X}_{+}$, $V\widehat{Y}_{+}$) and the
difference ($V\widehat{X}_{-}$, $V\widehat{Y}_{-}$) variances of the
two quadratures:

\begin{equation}
V\widehat{X}_{\pm }=\frac 12<(\delta \widehat{X}_1\pm \delta \widehat{X}%
_2)^2>,
\end{equation}

\begin{equation}
V\widehat{Y}_{\pm }=\frac 12<(\delta \widehat{Y}_1\pm \delta \widehat{Y}%
_2)^2>,
\end{equation}
where $\delta \widehat{X}_{1(2)}$ and $\delta \widehat{Y}_{1(2)}$
are the fluctuations of amplitude ($\widehat{X}$) and phase
($\widehat{Y}$) quadratures for the optical beam 1 (2),
respectively. If beam 1 and beam 2
are the coherent light with the noise at QNL, we have $V\widehat{X}%
_{+}^{coh}=V\widehat{X}_{-}^{coh}=V\widehat{Y}_{+}^{coh}=V\widehat{Y}%
_{-}^{coh}=1$ which are taken as the normalized QNL for measuring
the
quantum correlations\cite{Bra,Bac}, when $V\widehat{X}_{+}$($V\widehat{Y}%
_{+} $) is smaller than 1 the amplitude (phase) quadratures of the
two beams are quantum anticorrelated and if
$V\widehat{X}_{-}$($V\widehat{Y}_{-}$) falls below 1 they are
quantum correlated. Such correlations can not be observed in
classical light and can not be consistently accommodated in a
classical theory, so we say that they derive from the quantum
property of light field and essentially represent the pure quantum
correlations apart from classical correlations.

The CV entanglement of two bright optical beams is connected to the
non-local correlations between quantum uncertainties of the
conjugate variables, amplitude and phase quadratures. There have
been a lot of theoretical publications to discuss the criteria of
the optical entangled states with the CV correlations\cite
{Reid1,Reid2,Wer,Sim,Duan,Loo,Cof,Ade1,Ade2}. For the experiments
where the observables are the amplitude and phase quadratures of
optical fields, it is convenient to use the following the sufficient
non-separability criterion for any two-mode bipartite
state\cite{Sim,Duan}

\begin{equation}
V\widehat{X}_{\pm }+V\widehat{Y}_{\mp }<2.
\end{equation}
For a pair of optical beams with $V\widehat{X}_{+}$ $<1$ and $V\widehat{Y}%
_{-}<1$ we say it is an entangled quantum state with
quadrature-amplitude anticorrelation and quadrature -phase
correlation. If $V\widehat{X}_{-}$ $<1$ and $V\widehat{Y}_{+}<1$ we
have an entangled state with quadrature-amplitude correlation and
quadrature-phase anticorrelation. We
are not able to obtain an entangled state with $V\widehat{X}_{+}$ $<1$ and $V%
\widehat{Y}_{+}<1$ (or $V\widehat{X}_{-}$ $<1$ and
$V\widehat{Y}_{-}<1$) simultaneously since it violates the
uncertainty principle of quantum mechanics\cite{Bac,Reid1,Reid2}.

On the other hand, the interference phenomenon between light waves
directly characterizes the classical coherence of optical field,
which can be quantified by the visibility (Vis) of the interference
fringes. For the interference between two optical beams with
different frequencies $\nu _1$ and $\nu _2$, we easily deduce

\begin{equation}
V_{is}=\frac{I_{\max }-I_{\min }}{I_{\max }+I_{\min
}}=Exp[-\frac{(\nu _1-\nu _2)^2}{2\Delta \nu ^2}],
\end{equation}
where, $I_{max}$ and $I_{min}$ are the maximum and the minimum
values of the interference intensities, respectively. In Eq.(4) we
have assumed that the two optical beams are produced from a light
source simultaneously and the delay time of the interference between
them is taken as zero. The $\Delta \nu $ is the Gaussian frequency
spread of each beam.

\begin{figure}
\centerline{
\includegraphics[width=3in]{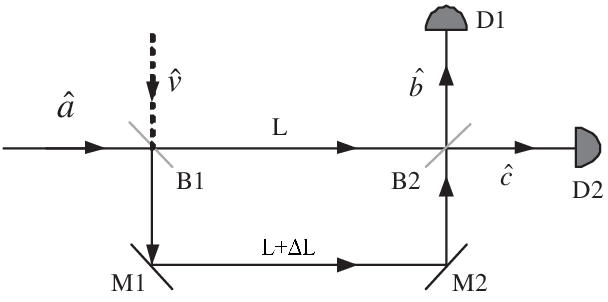}
} \vspace{0.1in}
\caption{The diagram of M-Z interferometer with an unbalanced arm
length. $B_{1(2)}$, 50\% beam-splitter;  $a$, input optical mode;
$v,$ the input vacuum field, $b$ and $c$,  two output modes.
\label{Fig1} }
\end{figure}

O. Gl\"{o}ckl et al. presented an elegant scheme which allow us to
perform the sub-shot-noise measurements of the phase and amplitude
quadratures of a bright optical beam as well as to determine the
corresponding QNL experimentally without the need of a separate
local oscillator by using a set of Mach-Zehnder interferometer with
unbalanced arm lengths\cite{Gl}.The diagram of the interferometer is
shown in Fig.1. $B_{1(2)}$ is a 50\% beam-splitter and $M_{1(2)}$ is
a mirror of 100\% reflectivity. $\widehat{a}$ is the measured input
optical mode, $\widehat{b}$ and $\widehat{c}$ are the two output
modes. $\widehat{v}$ denotes the vacuum field with a vacuum noise
$\delta \widehat{X}_0$ $=\delta \widehat{Y}_0=1$ and a zero average
intensity. $D_1$ and $D_2$ are the photodiodes to detect the photocurrents. $%
L$ and $L+\Delta L$ are the lengthes of the short and long arms of
the interferometer, respectively. It has been demonstrated when the
relative optical phase shift between two optical fields on $B_2$
equals to $\varphi =\pi /2+2k\pi $ (k an integer) and the phase
shift of the spectral component of rf (radio frequency) fluctuations
at a sideband mode ($\Omega $) is controlled to $\theta =\pi $, the
fluctuations of the sum and the difference photocurrents of
$\widehat{b}$ and $\widehat{c}$ in the frequency space are
proportional to the vacuum noise level [$\delta \widehat{X}_0(\Omega
)$] and
the spectral component of the phase quadrature of mode $\widehat{a}$ [$%
\delta \widehat{Y}_a(\Omega )$], respectively. That is, the
evaluations of the sum [$\widehat{n}_b(\Omega )+\widehat{n}_c(\Omega
)$] and the difference [$\widehat{n}_b(\Omega )-\widehat{n}_c(\Omega
)$] of the spectral components at sideband frequency $\Omega $
are\cite{Gl,Su}

\begin{equation}
\delta \widehat{n}_b(\Omega )+\delta \widehat{n}_c(\Omega )=a\delta \widehat{%
X}_0(\Omega ),
\end{equation}

\begin{equation}
\delta \widehat{n}_b(\Omega )-\delta \widehat{n}_c(\Omega )=a\delta \widehat{%
Y}_a(\Omega ),
\end{equation}
where $\widehat{n}_b=\widehat{b}^{+}\widehat{b}$ and $\widehat{n}_c=\widehat{%
c}^{+}\widehat{c}$ are the photon number operators of the output modes $%
\widehat{b}$ and $\widehat{c}$, $a$ is the classical amplitude ($a$
is
assumed to be real) of the input mode $\widehat{a}$. $\delta \widehat{X}%
^{coh}(\Omega )=a\delta \widehat{X}_0(\Omega )=a$ is the normalized
QNL of a coherent light with a classical amplitude $a$. In the
experimental measurements the value of $a$ always is taken as 1 for
simplification and without losing the generality. Removing $B_1$ the
mode $\widehat{a}$ directly arrives $B_2$ and the two detectors
($D_1$ and $D_2$) constitute a normal balanced detection system. In
this case we have\cite{Gl,Su}

\begin{equation}
\delta \widehat{n}_b(\Omega )+\delta \widehat{n}_c(\Omega )=a\delta \widehat{%
X}_a(\Omega ),
\end{equation}

\begin{equation}
\delta \widehat{n}_b(\Omega )-\delta \widehat{n}_c(\Omega )=a\delta \widehat{%
X}_0(\Omega ).
\end{equation}
It means that the sum (Eq.7) and the difference (Eq.8) photocurrents
evaluate the fluctuation of the amplitude quadratures and the QNL of
the input mode $\widehat{a}$, respectively. Therefore using the
unbalanced M-Z interferometer we can conveniently measure the
quantum fluctuations of the amplitude and the phase quadratures as
well as scale the QNL of an optical beam.

\begin{figure}
\centerline{
\includegraphics[width=3in]{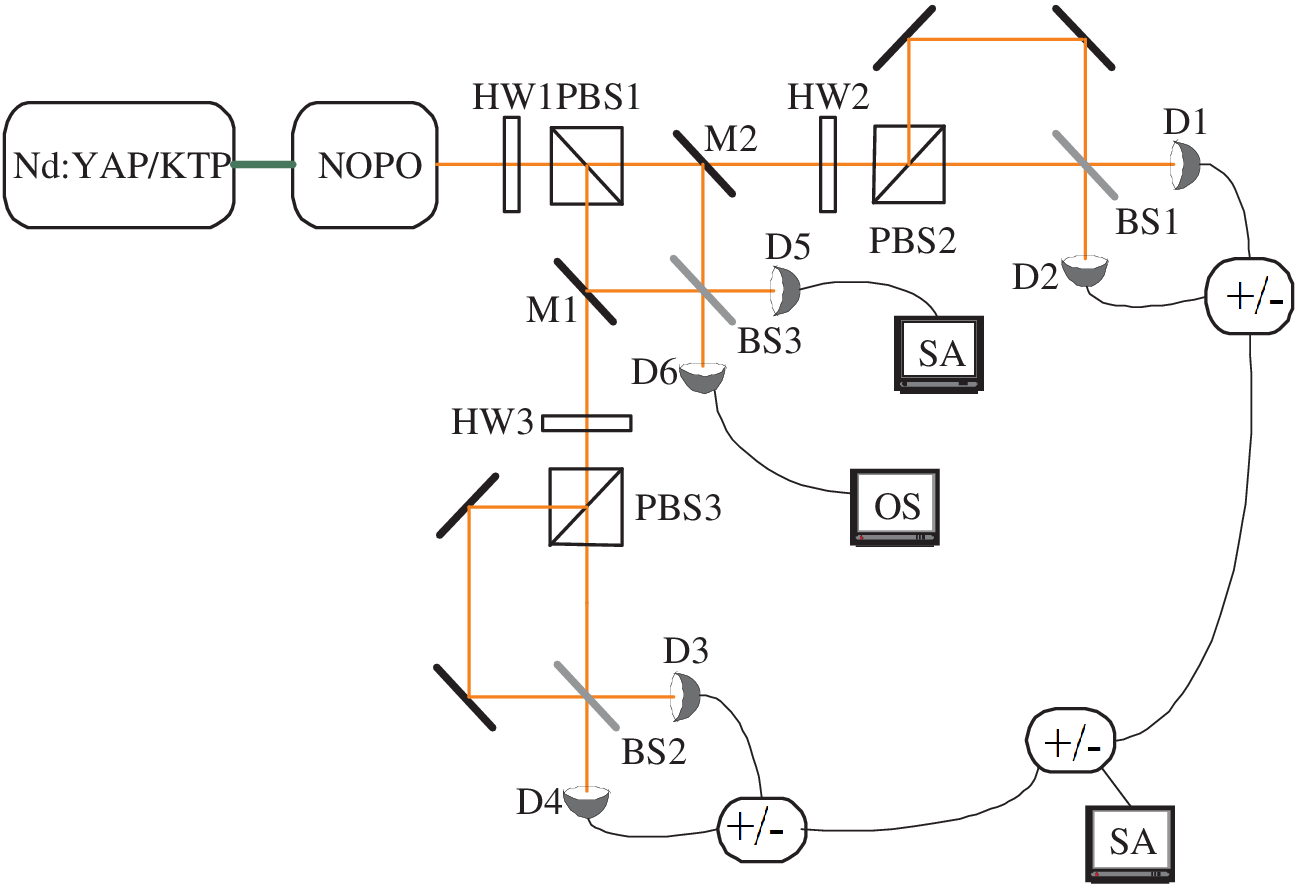}
} \vspace{0.1in}
\caption{(Color online) Schematic of experimental setup: Nd:YAP/KTP,
laser source; HW1-3, $\lambda /2$ wave plate; PBS1-3, polarizing
beam splitter; BS1-3, 50/50 beam splitter; M1-2, movable reflection
mirror; D1-6, ETX500 InGaAs photodiode detectors; SA, spectrum
analyser; OS, oscilloscope. \label{Fig2} }
\end{figure}

The experimental setup is depicted in Fig.2. The bright optical twin
beams with different frequencies are generated from a nondegenerate
OPO (NOPO) consisting of a type-II KTP (KTiOPO$_4$) crystal inside
an optical resonator, which is pumped by an intracavity
frequency-doubling Nd:YAP (Nd-dropped YAlO$_3$) laser. The
second-harmonic wave at 540 nm from the laser serves as the pump
field of the NOPO. Through a frequency down-conversion process
inside the NOPO above the threshold a pair of bright optical twin
beams with orthogonal polarizations is obtained in the output field,
which just is the signal and idler modes in the KTP
crystal\cite{Li,Reid1,Reid2}. The output twin beams are separated by
a polarizing-beam-splitter (PBS1) and then they are detected by the
two sets of unbalanced Mach-Zehnder (M-Z) interferometer,
respectively. The polarizing-beam-splitter PBS2 and PBS3 serve as
the input beam-splitters (corresponding to B1 in Fig.1) of the two
interferometers, respectively. The three half-wave plates HW1, HW2
and HW3 are used for the polarization alignment of the input beams
on the three PBSs respectively. By rotating the polarization
orientation of HW2 and HW3 we can conveniently switch between
amplitude and phase quadrature measurements. The beam-splitter BS1
(BS2) of 50\% is the output coupler of the interferometer, the
output beams of which are detected by the photodiodes D1 and D2 (D3
and D4). In the two arms of the interferometer the light beams
transmit in the optical fibers with a refraction of 1.55. The length
difference $\Delta L$
between the long and the short arm is 48m to satisfy the requirement of $%
\theta =\pi $ for the phase-quadrature measurement at the sideband
frequency of $2MHz$\cite{Gl}. The measured photocurrents by D1 and
D2 (D3 and D4) are combined with the positive or negative power
combiner (+/-) to give the QNL and the quadrature fluctuations
according to Eqs.(5)-(8). When two movable mirrors M1 and M2 of
100\% reflectivity are moved respectively in one of the twin beams,
the reflected beams from each mirror are combined on a beam-splitter
of 50\% (BS3) for the measurement of classical coherence. The output
photocurrents from the detector D5 is connected to a spectrum
analyzer (SA) for measuring the frequency difference between twin beams ($%
|\nu _1-\nu _2|$) by means of their beating signal. The output from
D6 is sent to an oscilloscope (OS) for measuring the visibility of
the interference fringes. Since the $\alpha $-cut KTP crystal used
in our system has a broad full temperature-width of about $40^OC$
around $63^OC$ for achieving the type-II noncritical phase matching,
we can tune the frequencies of twin beams by changing the
temperature of KTP which is placed in an electronic
temperature-controlled oven.

\begin{figure}
\centerline{
\includegraphics[width=3in]{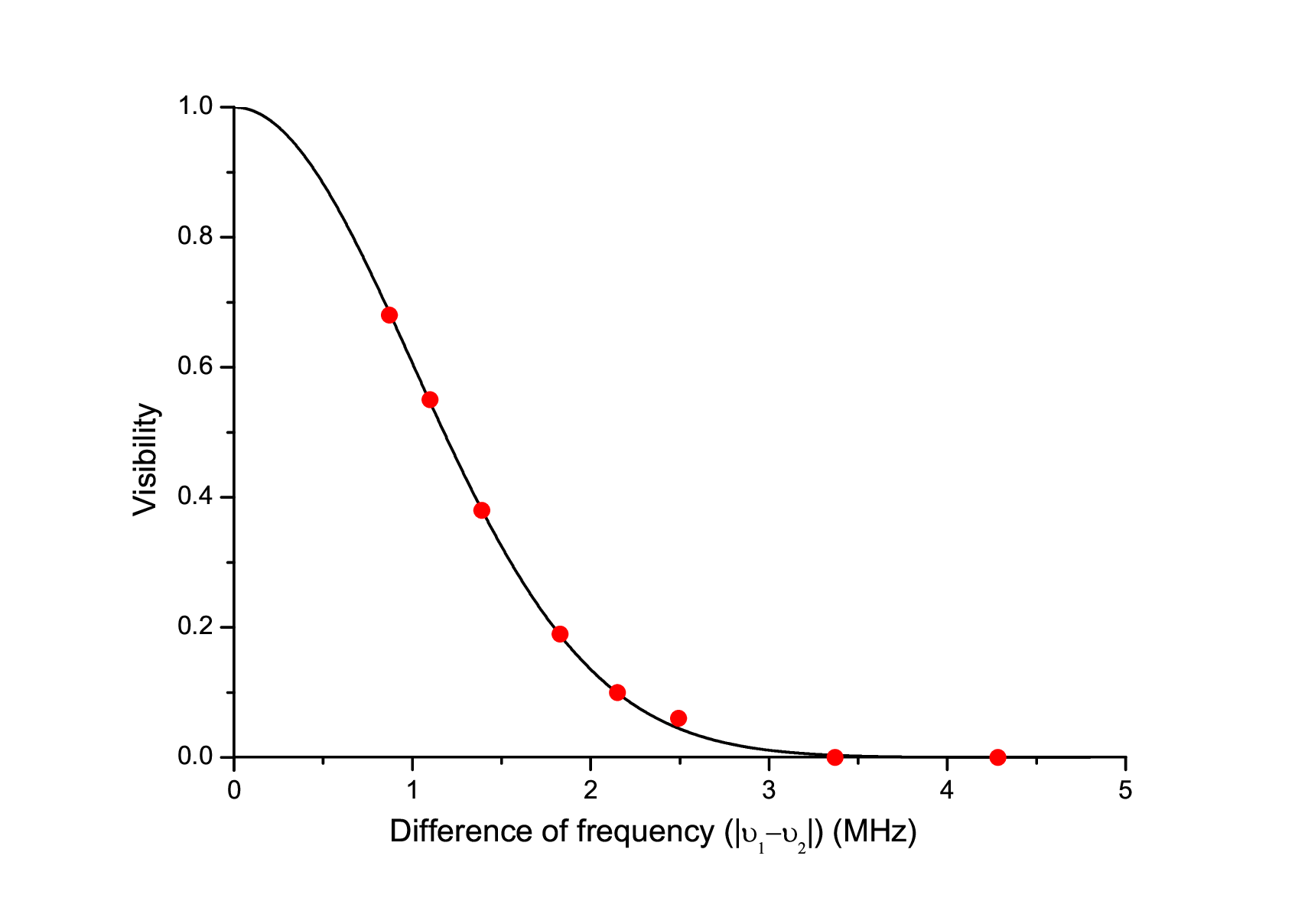}
} \vspace{0.1in}
\caption{(Color online) The dependence of the visibility of
interference fringes on the difference of frequency ($|\nu _1-\nu
_2|$). Solid line is the calculated curve with Eq.4 ($\Delta \nu
=1.0MHz$); $\bullet $, the measured data. \label{Fig3} }
\end{figure}

Fig.3 shows the dependence of the visibility of the interference
fringes on the frequency difference ($|\nu _1-\nu _2|$). The
frequency spread of each beam from the NOPO is $\Delta \nu $
$\approx 1.0MHz$. The solid curve calculated with Eq.4 is in good
agreement with the experimentally measured data ($\bullet $). When
($|\nu _1-\nu _2|$) increases to $1.41MHz$ the visibility reduces to
$1/e$ and when $|\nu _1-\nu _2|>3.37MHz$ the interference fringes
totally vanish ($V_{is}\sim 0$). In this case, we say, there is no
any classical coherence between the two optical beams.

\begin{figure}
\centerline{
\includegraphics[width=3in]{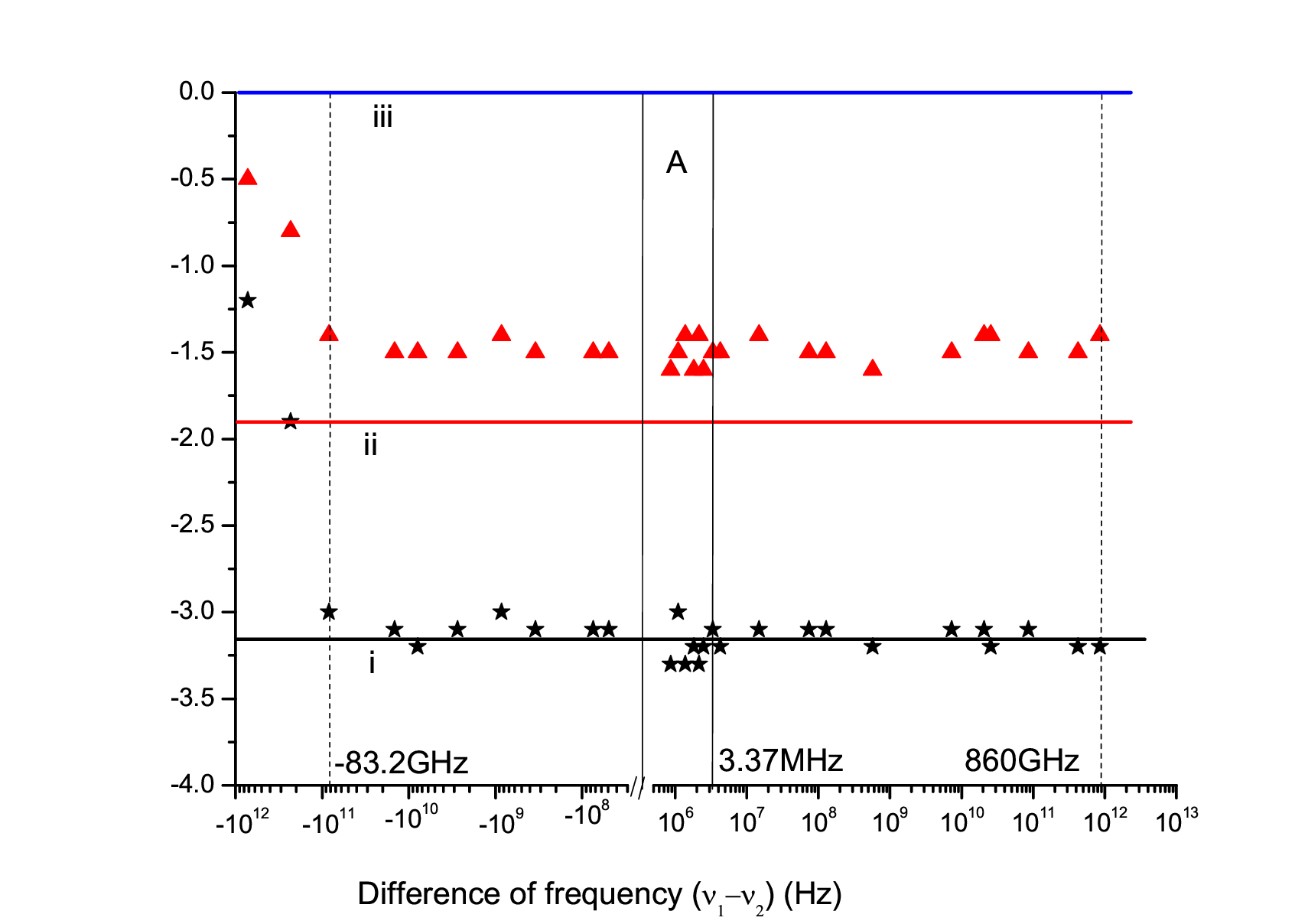}
} \vspace{0.1in}
\caption{(Color online) The measured noise power of the twin beams
at 2 MHz dependent on the difference of frequency $(\nu _1-\nu _2)$.
$i$, the calculated of amplitude correlation variance; $ii$, the
calculated of phase anticorrelation variance; $iii$, the QNL;
$\bigstar $, the measured amplitude correlation variances;
$\blacktriangle $, the measured phase anticorrelation variances. In
the region A of $(|\nu _1-\nu _2|)<3.37MHz$ both classical and
quantum correlations exist. Outside the region A only quantum
correltions exist without classical coherence. \label{Fig4} }
\end{figure}

Removing M1 and M2, we measured the quantum correlations of the twin
beams. Subtracting (adding) the two photocurrents of the amplitude
(phase) quadratures measured respectively by each M-Z interferometer
according to Eq.7 (Eq.6) with a negative (positive) power combiner
(+/-), we can
obtain the correlation (anticorrelation) variances $V\widehat{X}_{-}$ ($V%
\widehat{Y}_{+}$) of the amplitude (phase) quadratures between the
twin
beams. The measured correlation variances are shown in Fig.4. Both measured $%
V\widehat{X}_{-}$ ($\bigstar $) and $V\widehat{Y}_{+}$
($\blacktriangle $) are below the normalized QNL (0dB) and the
correlations almost keep constant
in the region of $-83.2GHz<(\nu _1-\nu _2)<860GHz$, where $V\widehat{X}_{-}$ and $V%
\widehat{Y}_{+}$ are $3.1\pm 0.1dB$ and $1.5\pm 0.1dB$ below the
QNL, respectively, which are in reasonable agreement with the
calculated results (solid line) according to the quantum correlation
formula deduced for NOPO above the threshold by Fabre et
al.\cite{Fab}, which are

\begin{equation}
V\widehat{X}_{-}=S_0[1-\frac{\eta \zeta ^2\xi }{1+(\frac fB)^2}],
\end{equation}

\begin{equation}
V\widehat{Y}_{+}=S_0[1-\frac{\eta \zeta ^2\xi }{\sigma ^2+(\frac
fB)^2}],
\end{equation}
where, $f=\Omega /2\pi $ is the noise frequency, $S_0$ is the QNL, $B$ and $%
\xi $ are cavity bandwidth and the output coupling efficiency of NOPO, $%
\zeta $ is the transmission efficiency of M-Z interferometer, $\eta
$ is the detective efficiency, and $\sigma $ ($=\sqrt{\frac
P{P_0}}$) is the pump parameter ($P$ is the pump power and $P_0$ is
the threshold pump power of
the NOPO). For our experimental system, the parameters are $\eta =90\%$, $%
\zeta =81\%$, $\xi =88\%$,$\ \sigma =\sqrt{\frac{195mW}{130mW}}=1.22$, $%
B=15.4MHz$ and $f=2MHz$. In the measurements of the correlation
variances (Fig.4) we use the normalized QNL ($S_0=1$, line $iii$ in
Fig.4). The measured amplitude correlation variances ($\bigstar $)
match perfectly with the calculated result from Eq.(9) (line $i$).
However the measured phase correlation variances ($\blacktriangle $)
are higher by $0.3dB$ than the theoretical calculation because the
worse mode-matching efficiency and the influences of the excess and
spurious phase noises in the pump field have not been involved in
the calculation with Eq.10 (line $ii$). The complicated factors
influencing the phase correlation of twin beams have
been analyzed detailedly in Refs.[24] and [25]. The measured value of ($V%
\widehat{X}_{-}+V\widehat{Y}_{+}$) is

\begin{equation}
V\widehat{X}_{-}+V\widehat{Y}_{+}=1.20<2.
\end{equation}
The result demonstrates the quantum entanglement of the twin beams
without the classical coherence.

For the conclusion, the presented experiment proves when the
frequency difference ($|\nu _1-\nu _2|$) between twin beams is
larger than $3.37MHz$, the interference fringes depending on the
classical coherence totally vanish, but the quadrature quantum
correlations are not influenced completely in the region of
$860GHz$. In the experiment, the frequency changes of the twin beams
were achieved by tuning the temperature of KTP crystal around the
central temperature ($\sim 51^OC$) with a degenerate frequency of
twin beams. When $(\nu _1-\nu _2)<-83.2GHz$ or $(\nu _1-\nu
_2)>975GHz$, the phase-matching condition for the nonlinear
interaction is broken down because the temperature of the crystal
has surpassed the phase-matching bandwidth. In this case the quantum
correlation will decrease
due to that the effective nonlinear coefficient of the KTP crystal is reduced%
\cite{Bra,Reid1}, thus the phenomenon of the correlation decrease
over phase-matching temperature region does not connect with the
classical coherence of light (Seeing the frequency region of $(\nu
_1-\nu _2)<-83.2GHz$ in Fig.4. The data in the region of $(\nu
_1-\nu _2)>860GHz$ were not measured since the temperature of the
KTP crystal is not able to be increased continuously due to the
limitation of the oven.).

In the quantum optics, the QNL is a detectable boundary between the
classical and the quantum effects. Using the boundary we verified
that the existence of the quantum correlations may totally not
depend on the classical coherence. The experiment provides us a
convenient system and scheme to generate and study the pure quantum
correlations with vanishing classical correlations. On the other
hand, it makes us sure that the quantum correlations can exist in a
pair of optical twin beams with very different frequencies, which is
important for quantum communication since one of them may match with
the atomic transition for information memory and other one may match
the transmission in optic fiber.

Acknowledgement: This research was supported by Natural Science
Foundation of China (Grants Nos. 60736040 and 60608012), NSFC
Project for Excellent Research Team (Grant No. 60821004), National
Basic Research Program of China (Grant No. 2006CB921101).

\end{document}